\documentstyle[aaspp4,10pt,epsf]{article}
\def\sun{\odot}
\def\lsim{\, \lower2truept\hbox{${< \atop\hbox{\raise4truept\hbox{$\sim$}}}$}\,}
\def\gsim{\, \lower2truept\hbox{${> \atop\hbox{\raise4truept\hbox{$\sim$}}}$}\,}

\def\oneskip{\vskip\baselineskip}

\def\oneskip{\vskip\baselineskip}

\begin{document}

\title{Mapping IR Enhancements in Closely Interacting Spiral-Spiral Pairs.
I. ISO~CAM and ISO~SWS Observations\footnote{Based on observations made 
with ISO, an ESA project with instruments
funded by ESA Member States and with the participation of ISAS and NASA.}}
\author{Cong Xu}
    \affil{Infrared Processing and Analysis Center, 
Jet Propulsion Laboratory, 
    Caltech 100-22, Pasadena, CA 91125;}
    \affil{Max-Planck-Institut f\"ur Kernphysik,
    Postfach 103980, D69117 Heidelberg, Germany}
\author{Yu Gao, Joseph Mazzarella, Nanyao Lu}
    \affil{Infrared Processing and Analysis Center, Jet Propulsion Laboratory, 
    Caltech 100-22, Pasadena, CA 91125}
\author{Jack W. Sulentic, Donovan L. Domingue}
   \affil{Department of Physics and Astronomy, University of Alabama,
           Tuscaloosa, AL 35487}
\received{February 28, 2000}
\accepted{ }

\begin{abstract}
Mid-infrared (MIR) imaging and spectroscopic observations are presented
for a well defined sample of eight closely interacting (CLO) pairs of
spiral galaxies that have overlapping disks and show enhanced
far-infrared (FIR) emission. The goal is to study the star formation
distribution in CLO pairs, with special emphasis on the role of
'overlap starbursts'. Observations were made with the Infrared Space
Observatory (ISO) using the CAM and SWS instruments. The ISO~CAM maps,
tracing the MIR emission of warm dust heated by young massive stars,
are compared to new ground based H$\alpha$ and R-band images.  We
identify three possible subgroups in the sample, classified according
to the star formation morphology: (1) advanced mergers (Arp~157,
Arp~244 and Arp~299), (2) severely disturbed systems (Arp~81 and
Arp~278), and (3) less disturbed systems (Arp~276, KPG~347 and
KPG~426). Localized starbursts are detected in the overlap regions in
all five pairs of subgroups (1) and (2), suggesting that they
are a common property in colliding systems. Except
for Arp~244, the 'overlap starburst' is usually fainter than the major
nuclear starburst in CLO pairs.  Star formation in 'less disturbed
systems' is often distributed throughout the disks of both galaxies
with no 'overlap starburst' detected in any of them. These systems also
show less enhanced FIR emission, suggesting that they are in an earlier
interaction stage than pairs of the other two subgroups where the direct
disk collisions have probably not yet occurred.
\end{abstract}


\section{Introduction}
It is well established from IRAS studies that many interacting galaxies
show enhanced far-infrared (FIR) emission compared to non-interacting
galaxies (Lonsdale et al. 1984; Kennicutt et al. 1987; Telesco et al.
1988). Most of the emission is due to young massive stars formed in
recent starbursts, supporting the idea that galaxy-galaxy interactions
can stimulate active star formation (Larson \& Tinsley 1978; Rieke et
al. 1980; Condon et al. 1982; Balzano 1983; Gehrz et al. 1983; Joseph
et al. 1984; Cutri \& McAlary 1985); albeit, dust-obscured AGNs might
be responsible for the emission in some cases (Sanders et al. 1988a;
Surace \& Sanders 1999; Genzel et al. 1998; Lutz et al. 1998). The most
extreme FIR luminosities ($\gsim 10^{12}\; L_\sun$; H$_0$=75 km/s
Mpc$^{-1}$) and highest dust temperatures are found in galaxy mergers
where the identity of the component galaxies is often indeterminate
(Sanders et al. 1988a; Melnick \& Mirabel
1990; Mazzarella et al. 1991;
Sanders \& Mirabel 1996). After these ultraluminous IR galaxies
(ULIRGs), violent starbursts are most likely to be found in closely
interacting pairs of spiral galaxies with overlapping disks such as
Arp~244 (the Antennae). These pairs dominate the FIR luminosity
function between $10^{11}$ --- $10^{12}\; L_\sun$ (Xu \& Sulentic
1991).

There is clear evidence in the recent deep surveys carried out at:  1)
optical/UV/NIR wavelengths by HST (Williams et al. 1996; Williams et
al. 1998; Thompson et al. 1999), 2) MIR/FIR wavelengths by ISO (Elbaz
et al. 1999; Puget et al. 1999), and 3) submm wavelengths by SCUBA
(Blain et al. 1999), that much of the evolution in the history of star
formation in the Universe is related to starbursts in closely
interacting/merging systems.  Therefore it is of great interest to
understand {\it how} the enhanced star formation (starburst) is
stimulated in these system.

The question of {\it how} starbursts are stimulated in interacting
galaxies, leads directly to the more fundamental question of how the
star formation is modulated in galaxies (Kennicutt 1989, 1998; Hunter
et al. 1998; Wyse \& Silk 1987; Dopita 1985). Since stars are formed
from gas, an obvious necessary condition for a galaxy to be
star-formation active is that it should contain significant gas. On the
other hand, it appears that interacting galaxies have in general higher
star formation rates (SFRs) not because they have more gas than
isolated galaxies, but because the star formation rate per unit gas
mass (the so called 'star formation efficiency') is much higher (Young
et al. 1986, 1989; Solomon \& Sage 1988; Sanders et al. 1991). As
pointed out by Kennicutt (1998) in his discussion of the 'Schmidt law'
(SFR$\propto \sigma_g^N$, N$\sim 1.4$, where $\sigma_g$ is the gas
surface density), the higher efficiency in starbursts in interacting
galaxies may simply be a consequence of their much higher gas densities
(Scoville et al.  1994; Solomon et al. 1997; Gao \& Solomon 1999;
Bryant \& Scoville 1999).  Unusually high gas density, which occurs
almost exclusively in the nuclei of interacting galaxies and mergers,
is certainly an interaction related phenomenon. Interaction induced gas
inflow is predicted by simulations and is  due either to the higher gas
viscosity (caused by more frequent cloud-cloud collisions in
interacting galaxies: Olson \& Kwan 1991a; Struck 1997), or to the
torque imposed on the gas by an interaction induced stellar-bar (Barnes
\& Hernquist 1996).

If enhanced star formation in interacting galaxies is due to higher gas
density, which has a simple physical interpretation related to the
growth rate of gravitational perturbation (Kennicutt 1998), two
predictions can be made:  
\begin{enumerate} 
\item Interaction induced starbursts should concentrate in the nuclear 
region.  
\item Starbursts in interacting systems of later stages (before the 
majority of the gas is converted to stars) should be stronger than those 
in early stages, because the later the stage, the more gas will sink into 
the nuclei.
\end{enumerate}
As far as the ULIRGs are concerned, these two predictions seem to be in
good agreement with the observations (see Sanders \& Mirabel 1996
for a review).
All ULIRGs appear to be in or close to the final stage of the
merger process (e.g. Sanders et al. 1991; Murphy et a. 1996;
Clements et al. 1996;
Clements and Baker 1996; Mihos \& Bothun 1998). The starbursts they
harbor are primarily in the nuclear regions. Mihos \& Bothun (1998)
noted a trend between dynamical age and $H\alpha$ concentration in the
ULIRGs they observed which is consistent with the physical scenario
behind the two predictions.

Statistical studies of a large pair sample (Xu \& Sulentic (1991)
demonstrated that a subsample of closely interacting
spiral-spiral pairs (hereafter CLO SS) with: a)
separations less than a component diameter and b) showing optical signs
of interaction, exhibit higher mean FIR--to--optical
luminosity and FIR color ratios. This FIR excess exists not only
with respect to isolated galaxies but relative to other pairs as well
(see also Telesco et al. 1988; Jones \& Stein 1989; Mazzarella et al.
1991). This result is again in line with prediction (2),
because the small separations (normalized to the primary component
diameter) of CLO SS pairs suggests that they may often be in a more
advanced evolutionary state of a merger sequence compared to other SS
pairs (Hwang et al. 1999). 
It should be noted, though, that: 1) since final coallescence usually
requires several encounters between a pair of galaxies (Barnes \&
Hernquest 1996, 1998) and 2) since the orbital geometry of interacting
systems is complicated, there is no one-to-one mapping between
component separation and interaction stage.  However, there is a clear
tendency for the apoapsis of the orbit to decrease rapidly after the
first encounter (Barnes \& Hernquest 1998). This suggests that {\it the
later the stage, the more chance for the two galaxies to remain close
together}.  On the other hand, the lack of one-to-one mapping between
component separation and interaction stage for individual galaxy pairs
(coupled with the projection effect) may be the reason for the absence
of a monotonic dependence between starburst strength and separation,
especially for pairs with wider separations (e.g. Fig.12 and Fig14 of
Xu \& Sulentic 1991; see also Keel 1993). 

Contradicting prediction (1), some famous examples of the CLO
SS pairs such as Arp~244 (Hummel \& van del Hulst 1986; Vigroux et
al. 1996; Mirabel et al. 1998) and Arp~299 (Gerhz et al. 1980; Hibbard
\& Yun 1998) show bright extranuclear starbursts in the overlap
region, possibly due to interpenetrating 
collisions between components (Jog \& Solomon 1992). Indeed, the simulation 
of Mihos et
al. (1993), which modeled star formation in Arp~244 with a
prescription similar to the 'Schmidt law', failed to reproduce the
overlap starburst. This motivates the following two questions:
\begin{enumerate}
\item Are overlap and other extranuclear starbursts common in CLO SS pairs?
\item If yes, then which starburst mode
is the more important in CLO SS pairs,
nuclear or overlap starbursts?
\end{enumerate}

One must map the star formation distribution in these systems with
spatial resolutions much better than IRAS in order to answer these
questions. On the other hand, since dust extinction in interacting
galaxies can be very large (Gehrz et al. 1983; Kunze et al. 1996),
optical star formation mapping may not provide reliable results.  These
considerations motivated us to map a sample of CLO SS pairs with
overlapping disks using two instruments (ISO~CAM and ISO~PHOT) on board
of the Infrared Space Observatory (Kessler et al. 1996). ISO-SWS
spectroscopic observations were also made for some pairs (pointing both
to the nuclei and to the overlap regions).  The imaging observations
provided higher angular resolutions than IRAS and orders of magnitude
higher sensitivity than KAO (e.g. Bushouse et al. 1998). This paper is
the first in a series that presents and analyzes the ISO observations.
We report the first results of ISO~CAM and ISO~SWS observations,
comparing them with results from new ground based H$\alpha$
observations. ISO~PHOT results and a more quantitative multiwavelength
analysis will be presented in a following paper.

\section{Observations}
\subsection{Sample Selection}
The pair sample was selected from two catalogues:
\begin{description}
\item{1.} Catalogue of Isolated Pairs of Galaxies in the Northern Sky 
(Dec.$ > -3$ deg)    by Karachentsev (1972), hereafter KPG;
\item{2.} Atlas of Peculiar Galaxies (Dec. $> -27$ deg) by Arp (1966).
\end{description}
The sample selection criteria were:
\begin{description}
\item{(1)} spiral--spiral pairs with overlapping disks;
\item{(2)} pair component redshift difference $\Delta$V$<$ 500 km s$^{-1}$;
\item{(3)} pairs showing one of the three interaction morphology classes 
defined in the KPG (LIN=bridges and/or tails, ATM=common halo, or 
DIS=distortion in one or both components);
\item{(4)} major axis diameter of the primary component D$> 1'$;
\item{(5)} a pair luminosity ratio $L_{fir}/L_B > 1$, where $L_{fir}$
is the integrated IRAS FIR (82.5 $\mu m$) luminosity (Helou et al. 1988) 
and $L_B$ is the combined monochromatic luminosity ($\nu L_\nu$) at 
4400{\AA} estimated from the photographic magnitudes.
\end{description}

Criterion (1) restricts the sample to galaxy pairs with overlapping
disks so that we can assess the role of overlap starbursts. This
differs from the criterion used to define CLO SS pairs in Xu \&
Sulentic (1991) which included systems with projected separations up
to  one primary component diameter.  Criterion (4) restricts the sample
to relatively nearby pairs and insuring that the chosen sample could be
resolved by ISO~PHOT (beamwidth $\theta$$\sim 45''$ at 60 and 100$\mu
m$).  The final criterion selects SS pairs with FIR/B $>$ 2$\times$ the
isolated spiral mean ($\langle L_{fir}/L_B \rangle = 0.5$, Xu \&
Sulentic 1991). This is also at least 2$\sigma$ above the mean $L_{fir}/L_B$
for normal spiral galaxies studied by Corbelli et al. (1991). Our
sample is biased towards FIR enhanced SS pairs in order to minimize the
chance of including false CLO SS pairs (wide pairs or accordant
chance alignments). It is worthwhile to mention that although FIR/B is
a good measure of starburst strength (Xu \& De Zotti 1989) it is
affected by several factors relatively unrelated to starburst activity
including B-band extinction and diffuse FIR cirrus emission.  Our
original sample contained ten CLO SS pairs but two (Arp~263 and  KPG~536)
were dropped because of poor ISO visibility parameters.  Basic
properties for our sample of eight CLO SS pairs are presented in Table
1.

\subsection{ISO~CAM Observations}
The 9.7$\mu m$ CAM images were obtained with the LW7 filter. They give
information about the (9.7$\mu m$) silicon feature and facilitate
comparisons with higher resolution ground-based (10$\mu m$) maps from
the literature (e.g. Gehrz et al. 1983; Bushouse et al. 1998). The 15
$\mu m$ CAM images were obtained with the LW3 filter except for the
very bright source Arp~299 (IRAS 12$\mu m$ flux of 3.7 Jy), where the
narrower LW9 filter (centered at 15$\mu m$ was used in order to avoid
saturation. These maps were expected to be significantly more sensitive
than the 9.7 $\mu m$ observations because the filter had a broader
bandpass and because the sources were expected to be brighter at longer
wavelength (less extinction and more emission). With a resolution of
$\sim 10^{\prime\prime}$ these CAM images reveal structures as small as
a few kpc in our sample where typical distances are  $\sim 40 Mpc$
(H$_0$=75 km/s Mpc$^{-1}$).  KPG~347 was a GTO target with the LW3
filter (Boselli et al.  1998).  Table 2 summarizes details of the
ISO~CAM observations.  Basic data reduction employed the CAM
Interactive Analysis (CIA) software\footnote{CIA is a joint development
by ESA Astrophysics Division and the ISO~CAM Consortium led by the
ISO~CAM PI, C. Cesarsky, Direction des Sciences de la Matiere, C.E.A.,
France.}.  Special attention was given to the correction of transient
effects using code based on the ``Fouks and Schubert algorithm''
developed by A. Abergel, A. Coulais, and H. Wozniak (Saclay) and
provided to us by M. Sauvage (1999, private communication).  This code
did a significantly better job than the standard CIA algorithms.

\subsection{ISO~SWS Observations}

We made spectroscopic observations of four emission lines (Br$_\beta$ -
2.63$\mu m$; Br$_\alpha$ - 4.05$\mu m$; [Ne~II] - 12.81$\mu m$ and
H$_2$ S(1) - 17.03$\mu m$) using ISO~SWS in the grating scan mode (AOT
SWS02).  Observations were obtained for three pairs in our sample
(marked by stars in Table 1) with three positions observed (galaxy
nuclei plus overlap region) in Arp~81 and 157 and two positions (both
nuclei) in Arp~278 (Table 3.1). Parameters of these observations are
given in Table 3.2. The purpose of these observations was to measure
star-formation rates and extinctions in different locations in the
pairs.  ISO~SWS reductions were performed using the SWS Interactive
Analysis package (IA3) developed by the international ISO SWS
Consortium.

\subsection{H$\alpha$ Observations}
H$\alpha$ and R-band observations were carried out at: 1) Palomar in
February 1996 using a Tek 1024$\times 1024$ CCD mounted on the 1.5m telescope
giving a scale of 0\farcs62 pixel$^{-1}$ and 2) Calar Alto in May 1996
and August 1997 using a 2048$\times 2048$ CCD on the 2.2 m telescope
giving a scale of 0\farcs33 pixel$^{-1}$ and/or 0\farcs53
pixel$^{-1}$.  Palomar narrow-band filters centered on redshifted
$H\alpha$ (+ [N~II]6548,6583) with FWHM $\sim 100$ {\AA} and adjacent
continuum broad-band filter (R-band) were used. Calar Alto narrow-band
(FWHM 50 -- 80 {\AA} centered on H$\alpha$) and R-band filters were
also employed.  The standard stars HD~84937, Landolt~104-334 (Palomar),
BD~+28~4211 and BD~+33~2642 Calar Alto were used for photometric
calibration. Standard IRAF data reduction procedures were used to
reduce this data.  Continuum subtracted H$\alpha$ images were produced
by scaling and subtracting the $R$ frames using field stars to match
the frames. Subtraction was carried out interactively until stellar
residuals were minimized.  Matching R band images were not obtained for
all Calar Alto targets.  In those cases continuum-free $H\alpha$ and
[N~II] emission line images were produced by subtracting the
off-centered adjacent narrow-band image from the one centered on
redshifted $H\alpha$ narrow-band maps, using about half dozen stars in
the field for normalization.  Aperture photometry was done using both
elliptical and circular apertures.

\section{Results}
\subsection{MIR Continuum v.s. H$\alpha$ Emission}
\subsubsection{Individual Pair Properties}
Figures 1--4 present  MIR and H$\alpha$ images for our sample. Three
images are presented for each source.  TOP PANEL: ISO~CAM 15$\mu m$
contours (9.7$\mu m$ for KPG~347) are plotted over an optical image
from the Digitized Sky Survey\footnote{The Digitized Sky Survey was
produced at the Space Telescope Science Institute under U.S. Government
grant NAG W-2166.  The images of these surveys are based on
photographic data obtained using the Oschin Schmidt Telescope on
Palomar Mountain and the UK Schmidt Telescope. The plates were
processed into the present compressed digital form with the permission
of these institutions. The complete acknowledgement can be found at
http://archive.stsci.edu/dss/dss\_acknowledgements.html.}.  Contour
levels are 2$^n$ (n=1,2,3,...) times the rms noise ($\sigma_{15\mu m}$
or $\sigma_{9.7\mu m}$ for KPG~347) as given in Table 2.  MIDDLE PANEL:
A log grayscale image of the 15$\mu m$ to $9.7\mu m$ MIR color ratio
with ISO~CAM 15$\mu m$ contours superposed (9.7$\mu m$ contours for
KPG347). Contour levels are 3+3$^{n}$ (n=0,1,2,3,...) times the rms
noise of the corresponding ISO~CAM image, with the range of the
grayscale image varied from source to source in order to maximize the
dynamic range. BOTTOM PANEL: H$\alpha$ contours plotted over the R-band
image.  Contour levels are different for each source because of the
different rms noise levels. 15 and 9.7$\mu m$ flux densities are given
for the sources in Table 4. The $f_{15\mu m}/f_{9.7\mu m}$ ratio
depends on both small grain emission and silicate absorption in the
9.7$\mu m$ band.  Detailed modeling of this ratio will be presented in
a future paper.

\vskip1truecm

\noindent\underbar{{\it Detailed notes}}

\noindent\underbar{Arp~299 = NGC~3690/IC~694}\footnote{Confusion exists 
over the name of the eastern component in Arp~299. We adopt IC~694 in
this paper. See Appendix in Hibbard \& Yun 1998 for details.}
\underbar{ (Figures 1abc)}: A pair of gas-rich galaxies IC~694 (east)
and NGC~3690 (west) regarded as a local example of a merger in progress
(Gehrz et al. 1983; Telesco et al.  1985; Wynn-Williams et al. 1991;
Casoli et al. 1999; Hibbard \& Yun 1999; Gallais et al. 1999).  Arp~299
is the most IR luminous system in our sample with $L_{fir}=
2.7~\times~10^{11}\; L_\sun$. Remarkably with $f_{60\mu m}=108.9$ Jy,
it is one of the brightest (60$\mu m$) extragalactic point sources in
the IRAS point source catalog (Soifer et al. 1987). The component disks
are in contact suggesting a relatively advanced stage of coallescence,
although the nuclei are well separated and are resolved by ISO~CAM. The
extranuclear starburst (Source C from Gehrz et al. 1983) is about
$8^{\prime\prime}$ north of the NGC~3690 nucleus (Source B in Gehrz et
al. 1983, the western galaxy) is only marginally resolved (seen as a
plateau on the contour plot) from the latter by ISO~CAM. Gallais et
al.  (1999) report ISO~CAM CVF observations of Arp~299 which cover
$48^{\prime\prime}\times 48^{\prime\prime}$ with pixel size of
$1.5^{\prime\prime}\times 1.5^{\prime\prime}$. This can be compared to
our observations (Table~2) which have a $3.8^{\prime}\times
3.8^{\prime}$ field of view and a $3^{\prime\prime}\times
3^{\prime\prime}$ pixel size.  Their LW3 (15$\mu m$) map clearly
resolves the overlap starburst (Source C) from the nucleus of
NGC~3690. 

The $f_{15\mu m}/f_{9.7\mu m}$ ratio map is generally smooth with
values close to the ratio of the integrated fluxes ($f_{15\mu
m}/f_{9.7\mu m}=2$: Table 4) across most of the disks. The nucleus of
IC~694 (Source A in Gehrz et al. 1983) and a few locations in the outer
disk show ratios as high as $f_{15\mu m}/f_{9.7\mu m}>10$. While high
values for the ratio may have large uncertainty in the outer regions,
the high ratio in the IC~694 nucleus implies a high silicate absorption
which may severely depress the 9.7$\mu m$ flux (see also Gehrz et al.
1983). This is consistent with results of CO imaging (Sargent \&
Scoville 1991; Aalto et al. 1997; Casoli et al.  1999) showing that the
nucleus of IC~694 has the highest surface density of molecular gas in
the Arp~299 system and comparable to values found in ULIRGs (Downes \&
Solomon 1998).  The inner part of the MIR emission region shows good
correspondence with the H$\alpha$ emission allowing for differences in
spatial resolution (Fig.~1, Panel c). This indicates that: 1) most of
the MIR emission is due to dust associated with star formation regions,
as suggested by Sauvage et al. (1996) in an ISO~CAM study of M51, and
2) the H$\alpha$ extinction is rather smooth, consistent with the
smooth $f_{15\mu m}/f_{9.7\mu m}$ ratio image.  The outer envelope of
the MIR emission is elongated in along PA$\sim$45$^{\circ}$ which is
also the ISO~CAM scan direction.  The elongation is contrary to that of
the H$\alpha$ emission suggesting that the ISO~CAM elongation is likely
to be an artifact of the transient behavior in the ISO~CAM detectors
(Cesarsky et al. 1996).  Careful transient corrections were included in
our reduction of this data, but the very high surface brightness (up to
$\sim 1$ Jy/beam) in the central region of Arp~299  produce strong
effects that are difficult or impossible to remove completely.
\vskip1cm

\noindent\underbar{Arp~244 = NGC~4038/9 (Figures 1def)}: 
The ``Antennae'' are regarded as another local example of a galaxy
merger (Rubin et al.  1970; Toomre \& Toomre 1972; Schweizer 1978;
Hummel \& van der Hulst 1986; Vigroux et al. 1996; Mirabel et al. 1998;
Evans et al. 1998; Whitmore et al. 1999). ISO~CAM images at 6.7 and
15$\mu m$ have been published (Vigroux et al. 1996; Mirabel et al.
1998). Our 9.7 and 15$\mu m$ maps cover a larger area ($6'\times 6'$).
The brightest region in this system at MIR wavelengths involves an
extranuclear starburst (Source A) about 15$^{\prime\prime}$ northeast
of the NGC~4039 nucleus (c.f. Vigroux. et al. 1996; Mirabel et al.
1998).  There is an H$\alpha$ peak at Source A, but much fainter than
the emission from the NGC~4039 nucleus. Apparently much of the optical
emission associated with Source A is extinguished by dust. More
intriguingly, the FIR (60$\mu m$, 100$\mu m$ and 160$\mu m$) KAO maps
(Evans et al. 1998; Bushouse et al. 1998) show that the FIR peak is
displaced north of Source A, at the position of a dark patch in the
optical images (e.g. the HST WFPC2 images by Whitmore et al. 1999). New
SCUBA maps (Haas et al. 2000) at 450 and 850$\mu m$
(15$^{\prime\prime}$ resolution) reveal large amounts ($10^{6-7}$
M$_\sun$) of cold dust ($<20$K) in the overlap region. There are
corresponding radio continuum peaks at both the MIR and FIR peaks
(Hummel \& van der Hulst 1986). The CO observations (Stanford et al.
1990; Gao et al. 1998; Gruendl et al. 1998) demonstrate that most of
the molecular gas in this system is extended throughout the overlap
region. The [C~II] 158$\mu m$ emission also peaks at the dark patch
north of Source A (Nikola et al. 1998). All of these observations
indicate that much of the star formation activity in Arp~244 is hidden
by dust. Our MIR maps also show some weak emission at the beginning of
the southern tail starts and a similar feature is observed in the radio
continuum (Hummel \& van der Hulst 1986).

The $f_{15\mu m}/f_{9.7\mu m}$ ratio map also peaks around Source A
with a value of $(f_{15\mu m}/f_{9.7\mu m})\sim 3$. Given that most of
the molecular gas is found near this region (Gao et al. 1998), and very
strong extinction (A$_V\simeq 70$) is found from an SWS study of MIR
lines in the same region (Kunze et al. 1996), the high $(f_{15\mu
m}/f_{9.7\mu m})$ level is likely due to significant silicate
absorption. 

In a separate ISO~CAM observation (Arp~244-02, Table~2),
we obtained ISO~CAM LW3 images further down along the southern
tail, including the end of the tail where a dwarf galaxy was found
(Schweizer 1978; Mirabel et al. 1992). No MIR emission is detected in
these tidal features.

\noindent\underbar{Arp~157 = NGC~520a/b (Figures 2abc)}: This is a 
very complex system apparently involving two (colliding) disk galaxies,
one oriented southeast-northwest (NGC~520a) and another oriented
east-west (NGC~520b) (Stanford 1991; Stanford \& Balcells 1990, 1991;
Bernl\"ohr 1993ab). The collision center appears to lie near the
nucleus of NGC~520b where where the MIR emission peaks.  The CO
emission also concentrates at this position (Sanders et al 1988b).  The
H~I gas shows the the kinematic signature of a rotating disk with the
same orientation as NGC~520a, while the rotation center is clearly at
the nucleus of NGC~520b (Hibbard \& van Gorkom 1996). An interesting
possibility is that the H~I originally belonged to the former but has
been captured by the latter (which might have an order of magnitude
more mass than the former; Bernl\"ohr 1993b). Note also that both disks
show rotation axes nearly parallel to the axis of the orbital motion
(Bernl\"ohe 1993b), making the hypothesis for migration of the H~I gas
more reasonable. There is a second, much weaker peak in the MIR
emission associated with the nucleus of NGC~520a, where new millimeter
synthesis observations failed to detect any CO emission (Hibbard et al.
2000, private communication).  The H$\alpha$ map (see also Hibbard \&
van Gorkom 1996; and Young et al. 1988) shows very different morphology
from the MIR maps which is most likely due to dust extinction of the
former. Near the nucleus of NGC~520b, heavy dust lanes are visible in
optical images. The $(f_{15\mu m}/f_{9.7\mu m})$ ratio, as a rough
indicator of silicate absorption, also peaks there. A NIR K-band image
of Arp~157 (Bushouse \& Werner 1990) shows similar morphology to the
MIR images, indicating that the difference between the MIR and
optical/H$\alpha$ morphologies is not due to different angular
resolutions. This is consistent with results of Bernl\"ohr (1993b)
whose model predicts that both NGC~520a and NGC~520b have been
undergoing starbursts, with the starburst associated with NGC~520a
about 2-3 $10^8$ years older and a factor of $\sim 8$ fainter than the
starburst associated with NGC~520b.

\noindent\underbar{Arp~81 = NGC~6621/2 (Figures 2def)}: 
This system was included by Toomre (1978), along with Arp~157 and 244,
in proposed sequence of mergers. A recent HST WFPC B-band image (Keel,
private communication) shows that NGC~6622 (the southern galaxy) is
likely to be a high inclination S0 galaxy. H$\alpha$ and MIR emission
from the center of NGC~6622 indicates that some star formation is
occurring there but at much lower level than in the nuclear region of
NGC6621. The HST image shows a star formation region between the two
galaxies along with gas/dust features extend from NGC~6621 and near the
center of NGC~6622.  There starburst in the overlap region is
seen in both H$\alpha$ and MIR maps. The tidal tail is a marginal
(2$\sigma$) detection on ISO~CAM images. The nucleus of NGC~6622 and
the overlap starburst show rather low $f_{15\mu m}/f_{9.7\mu m}$ ratios
($\sim 1)$, while the nucleus of NGC~6621 shows a  higher level
$f_{15\mu m}/f_{9.7\mu m}$ ($\sim 3$).  A KAO 100$\mu m$ observation
(Bushouse et al. 1998) detects only the NGC~6621 nucleus. Given the
large beam ($\sim 50^{\prime\prime}$) it is possible that the overlap
starburst may have contributed significantly to the flux.  A 1.426 GHz
VLA map (Condon et al. 1996) peaks on the NGC~6621 nucleus with an
extension that may be due to the overlap starburst or even emission
associated with NGC~6622.

\noindent\underbar{Arp~278 = NGC~7253a/b (Figures 3abc)}: 
This system looks like a more edge-on version of the famous ``Taffy
Galaxies'' (UGC~12914/5) (Condon et al. 1993; Jarrett et al. 1999).
Both galaxies are highly inclined, and much diffuse emission is
observed between the two disks (as in case of ``Taffy'').  
Arp~278 shows evidence of starbursts
in both galaxies (Bernl\"ohr 1993a) with a possible time delay of
10-70 Myr between the components (the larger galaxy shows the younger
burst).  Both galaxies are detected in the radio continuum (Condon et
al.  1996).  MIR emission from the northwestern galaxy (NGC~7253a, the
primary) peaks at the nucleus, while the peak is offset (toward the
primary) in the other component.  From deep optical image, the two
galaxies appear to be in contact near the nucleus of the southeastern
galaxy (NGC~7253b), where several H$\alpha$ knots exist
(Fig.3c). Given the clear link between the star formation regions in
the two galaxies as shown in the H$\alpha$ image, at least some of the
emission in those H$\alpha$ knots is likely due to overlap starburst.
In connection with the Taffy Galaxy analogy (Condon et al. 1993;
Jarrett et al. 1999), the components of Arp~278 may have recently
undergone a face-on collision. The plumes (i.e. the diffuse emission
between the two disks) may be viewed as debris from that
collision. Keel (1993) suggests that NGC~7253a is undergoing a direct
(prograde) encounter. Given that NGC~7253a also dominates the MIR
emission of the pair (Table 4), it is likely to be the more disturbed
component that is the source of most of the debris.  The $(f_{15\mu
m}/f_{9.7\mu m})$ ratio map peaks at the nucleus of NGC~7253a
although, in general, that ratio is not high ($<2$) indicating that
the overall dust extinction is less severe than in Arp~244 and 299.

\noindent\underbar{Arp~276 = NGC~935/IC~1801 (Figures 3def)}: 
Neither galaxy in this pair shows evidence for strong perturbation.
Both the MIR and H$\alpha$ emissions extend over the galaxy disks.
Most of the H$\alpha$ emission in the primary comes from H~II regions
in the spiral structure while the nucleus shows little or no emission.
This $H\alpha$ ``hole'' might be due to over-subtraction of the continuum
or, alternatively, may indicate a high extinction in the nuclear region
where the MIR emission peaks. A foreground star is projected between
the two component nuclei (but not on the overlap region)  making
H$\alpha$ measures unreliable there.  H$\alpha$ emission from the
smaller galaxy appears more concentrated.  with the nucleus
contributing significantly to the total flux.  Two H~II regions on
either side of the nucleus and perpendicular to the major axis,
contribute much of the remaining H$\alpha$ flux.  No localized
H$\alpha$ or MIR feature is detected in the overlap region.  Given the
differences in angular resolution, the $f_{15\mu m}$ and H$\alpha$ maps
shows similar morphology (i.e. the emission in the primary is extended
and that in the secondary is more centrally concentrated.  The
$(f_{15\mu m}/f_{9.7\mu m})$ ratio map is smooth with little dependence
on MIR surface brightness. It is unlikely that dust extinction is high
in this pair and the temperature of small grains, which are the major
contributors to  $f_{9.7\mu m}$ and $f_{15\mu m}$, does not depend
sensitively on the  radiation intensity (D\'esert et al. 1990).

\noindent\underbar{KPG~347 = NGC~4567/8 (Figures 4abc)}: 
This pair in the Virgo cluster is known as the  ``Butterfly
Galaxies''.  The 9.7$\mu m$ contours are plotted in Fig.~4. The
$(f_{15\mu m}/f_{9.7\mu m})$ ratio image used a 15$\mu m$ map from
Boselli et al (1998).  The two galaxies in this system also  show
little morphological disturbance.  A VLA H~I map of the system (Cayatte
et al. 1994) shows that the H~I distribution of the component disks to
be reasonably intact. Star formation is widespread in both galaxies and
unlike the CLO SS pairs discussed above, the $(f_{15\mu m}/f_{9.7\mu
m})$ ratio shows dips rather than peaks at the positions of the nuclei
with high ratio plateaus in the outer disk. The reason for the
variations may be different in this pair, related perhaps to heating of
small grains that are responsible for MIR continuum emission. In more
active pairs (e.g. Arp~299)  silicate absorption is the more likely to
be the major cause of variations in the flux ratio.

\noindent\underbar{KPG~426 = UGC~9376a/b (Figures 4def)}: 
Deep optical image show that the component of this pair are embedded in
a common envelope. Both the MIR and H$\alpha$ morphologies look
undistorted (symmetric). Most of the star formation in the northern
galaxy is concentrated in the nucleus while that in the southern galaxy
shows a nuclear and ring structure. Detailed comparison between the
morphology and simulations (Toomre and Toomre 1972) suggest that the
northern galaxy has undergone a retrograde encounter, while the
southern galaxy a prograde but high inclination ($>30^\circ$)
encounter. Similar to KPG~347, the $(f_{15\mu m}/f_{9.7\mu m})$ ratio
shows dips rather than peaks in the nuclei.

\subsubsection{Collective Properties}
The eight systems studied in this paper provide a small but
representative sample of CLO SS pairs with enhanced FIR emission. The
MIR and H$\alpha$ observations provide direct information about the
{\it distribution} of  star formation, which may impose new constraints
on models for interaction induced star bursts. On the other hand, the
uncertainties associated with small sample statistics make the following
inferences suggestive rather than conclusive.

We divide our sample into three subgroups based upon the star formation 
morphology:
\begin{description}
\item{I.} Advanced mergers (Arp~157, 244 and 299): The nuclei in these
systems are still well separated (or they would not have been cataloged
as pairs by Arp and/or Karachentsev) however both the stellar and
gaseous disks are ``entangled'' with each other. These pairs have the
least separations ($<8 kpc$) in our sample. They also show line
of sight velocity differences $\Delta$V$\leq$60 km s$^{-1}$, 
consistent with a significant dissipation of orbital angular momentum.
Simulations (e.g.  Mihos et al. 1993; Hibbard \& Yun 1999) suggest that
these pairs will merge in a few times of 10$^8$ years.  
\item{II.} Severely disturbed systems (Arp 81 and 278): The 
two disks remain separated but severe tidal distortions are present.
Line of sight $\Delta$V$\sim$200 km s$^{-1}$ for these pairs.
\item{III.} Less disturbed systems (Arp~276, KPG~347 and 426): Galaxies in
these pairs show reasonably normal morphologies in both  MIR and
optical/H$\alpha$ images. Although all three pairs are classified ATM
or DIS by Karachentsev (1972), our R-band images show nearly normal
(Arp~276 and KPG~347) or only weakly distorted (KPG~426) morphologies
in contrast to subgroups 1) and 2).  $\Delta$V ranges from 50-150 km
s$^{-1}$ for these binary systems.
\end{description}

The following trends can be drawn from our observations:
\begin{description}
\item[1)] All five CLO pairs (Arp~81, 157, 244, 278 and 299) 
in the first two subgroups show
localized MIR and H$\alpha$ emission enhancements in the overlap region.
 Only in Arp~244 is the overlap region
starburst more luminous than the nuclei burst while in Arp~81 and 299
the overlap region starbursts are significantly fainter than the most
active nucleus in the system (in Arp~157 and 278 the overlap region is close to
one of the nuclei). The inference is that star formation
induced by the hydrodynamic collisions of gaseous disks is a common
phenomenon in closely interacting galaxy pairs. However, it is
apparently seldom the dominant source of starburst emission in these
systems. The nuclear starburst, driven presumably by the gravitational
tidal force which induces infall of large quantities of gas into galaxy
nuclei, is perhaps the major mechanism for the enhanced star formation in
these galaxy pairs.
\item[2)] CLO pairs in the third subgroup show star formation that is
more evenly distributed in the component galaxies. This includes both
disk and nuclear activity as is
observed in isolated samples at a lower average intensity level.  H~II
emission regions in these components closely follow the spiral arms,
similar to the distribution of star formation in M51 (Sauvage et al.
1996). No overlap starbursts are observed in these systems.  With a
caution about small sample statistics, they also show relatively low
FIR/B ratios ($<L_{fir}/L_B>=1.30\pm 0.25$) and cool FIR colors
($<f_{60\mu m}/f_{100\mu m}> = 0.36\pm 0.02$) compared to pairs in the
first two subgroups which show $<L_{fir}/L_B>=4.93\pm 1.37$ and
$<f_{60\mu m}/f_{100\mu m}> = 0.65\pm 0.09$.
\end{description}

\subsection{MIR Line Emission}
The in-orbit performance of SWS over the observed wavelength regions
was much lower than expected at the time our observations were
planned.  No Br$\gamma$ or Br$\alpha$ features were detected while
[N~II]~12.81$\mu m$ and H$_2$~S(1)~17.03$\mu m$ were only detected in
Arp~157.  The ISO~SWS results are presented in Table 6. The
[N~II]~12.81$\mu m$ line and H$_2$~S(1)~17.03$\mu m$ lines were only
detected in the nucleus of NGC~520b (Fig.~2ab). A pointing at the
overlap region in this pair also detected the [N~II]~12.81$\mu m$ line
(Fig.~2c) but this may be contaminated by emission from the NGC~520b
nucleus because of the large beam.  A Br$\gamma$ line (EW= 17.0$\pm
0.5$ {\AA}) was previously reported by Vanzi et al. (1998).

SWS observations of Arp~244 and 299 exist in the literature. The MIR
line ratios, assuming that gas and dust are well mixed, imply rather
high dust extinctions: A$_V \sim 20$ for Arp~299 and A$_V \sim 80$ for
Arp~244 (Kuntz et al. 1996; Genzel et al. 1998).  It is interesting to
compare our new data for Arp~157 with SWS observations of related
objects. The [N~II]~12.81$\mu m$ flux for NGC~520b is comparable to the
values detected in the SWS survey of ULIRGs as well as template
starburst and AGN sources (Genzel et al. 1998).  A more physically
meaningful comparison between sources can be made using the ratio of
the [N~II]~12.81$\mu m$ line to total FIR flux estimated from the IRAS
observations using $FIR = 1.26\times10^{-14}~(2.58f_{60\mu m} +
f_{100\mu m}) [W~m^{-2}]$ (e.g., Sanders \& Mirabel 1996). The ratio of
$f([N~II])/f(FIR) = 0.001$ for NGC~520b similar to the values observed
in NGC~6240 and Arp~244, and about half value observed for the nuclei
of Arp~299 in the more sensitive observations reported by Genzel et al.
(1998).  This is another direct indication that the physical conditions
of active star formation in NGC~520b are similar to those observed in
Arp~244 and 299, the other two sources in our first subgroup.

\section{Discussion}

\subsection{Overlap Starbursts}
What is special about the overlap region? It is the actual interface
region of a disk-disk collision. An overlap starburst could then be
triggered by direct collisions between giant molecular clouds (GMCs)
(Noguchi 1991). At the same time this mechanism may not be very
efficient because the galactic disk filling factor of GMCs is very low
($> 0.01$). Jog \& Solomon (1992) proposed a more efficient model where
collisions between H~I clouds lead to the formation of a hot ionized,
high-pressure remnant gas. The over-pressure due to this hot gas causes
a radiative shock compression of the outer layers of existing GMCs in
the overlap region. These layers become gravitationally unstable and
trigger a burst of massive star formation in the initially barely
stable GMCs. This model can be tested with X-ray observations of the
hot remnant gas.  High resolution ROSAT observations of Arp~244
(Fabbiano et al. 1997) revealed X-ray emission in the overlap region.
However, given the morphology of the X-ray emission (localized,
point-source like), it is more likely due to supernova remnants
associated with the on-going starburst rather than a hot remnant gas
which would be more diffuse. The X-ray emission and star formation
associated with the ongoing collision in Stephan's Quintet (Pietsch et
al. 1997; Xu et al. 1999; Sulentic et al. 2000) may be closer to this
situation.  Future higher resolution AXAF observations could provide
more definite results.

If cloud-cloud collisions can indeed trigger star formation (Scoville et
al. 1986; Olson \& Kwan 1991ab) then  this may be a mechanism (in
addition to the gas-density dependence of star formation rate) for
interaction induced star formation enhancements in general.
Simulations (Olson \& Kwan 1991ab; Noguchi \& Ishibashi 1986; Noguchi
1988, 1991) have shown that cloud-cloud collisions are significantly
enhanced throughout the disk due to the orbit-crossing of gas clouds
triggered by gravitational perturbations. This may provide an
interpretation for more widely distributed star formation, that shows
more moderate enhancement, in ``less disturbed'' CLO pairs.

It is interesting to compare the overlap starbursts in Arp~299 (Source
C-C' in Gehrz et al. 1983) and Arp~244 (Source A in Vigroux et al.
1996). Table 6 gives the MIR fluxes of overlap starbursts and galactic
nuclei for these two systems. The overlap starburst in Arp~244 is
significantly brighter than both nuclei combined!  On the other hand,
the overlap starburst in Arp~299 is much fainter than either of the
component nuclei.  One possible reason for this difference is that the
two pairs are in different stages of the merging process.  
Arp~299 may be in a later merging stage where more gas has
fallen into the galactic nuclei powering intensive starbursts
(especially in IC~694; Casoli et al. 1999). Interaction kinematics may
also play a role in Arp~299 where IC~694 is undergoing a retrograde
encounter (Augarde \& Lequeux 1985; Hibbard \& Yun 1999). This
apparently allows it to retain most of its gas which is then available
to the nucleus as fuel for the violent starburst. On the other hand,
both disks in Arp~244 are undergoing prograde encounters (Toomre \&
Toomre 1972; Mihos et al. 1993), and the gas disks have suffered severe
disruption (Barnes \& Hernquist 1996, 1998) with much of the gas moved
away from the central regions (van der Hulst 1979; Gao et al. 1998;
Grundl et al. 1998).  The highest gas concentration is, in fact, found
in the overlap region (Stanford et al. 1990; Gao et al. 1998; Grundl et
al. 1998).  It is possible that such a gas distribution is unstable and
transient (retardation of infall) which might be the reason 
why such bright overlap starbursts are rare.

\subsection{'Less Disturbed Systems'}
Our third CLO SS subgroup shows a modest FIR enhancement but less
structural distortion and more extended star formation.  These
properties are consistent with the hypothesis that they are in an
earlier interaction stage than the first two subgroups.  The absence of
overlap starbursts in these may indicate that a disk-disk collision has
not yet occurred.  Although no firm conclusion can be reached about
their physical separation, the calculation of Xu \& Sulentic (1991,
Appendix A) shows that components in most close pairs are physically
proximate (physical separation $\leq$ diameter of the primary).
Examination of component morphologies in KPG~426, the pair with
components most clearly separated in our sample, and consideration of
simulations (e.g. Toomre \& Toomre 1972), suggest that the components
are well separated. There is, at the same time, evidence that 'less
disturbed systems' may have unfavorable interaction geometries.  Two
out of three pairs in subgroup three may be undergoing either
retrograde or highly inclination encounters (the situation for Arp~276
is less clear).

\subsection{Variations in the Strength of Star Formation Activity} 

We suggest that the following sequence of sources of increasing
star formation efficiency: isolated galaxies -- wide pairs -- closely
interacting pairs -- ULIRGs, may be viewed as an evolutionary
sequence for mergers.  Results in this paper show
further that, within the population of closely interacting
pairs (CLO SS), the 'less disturbed systems' may be in earlier stages of
interaction and have less enhanced star formation activity compared
to 'severely disturbed systems' and 'advanced mergers'.

At the same time, CLO SS pairs show a large scatter in star formation
indicators such as $L_{fir}/L_B$ and $f_{60\mu m}/f_{100\mu m}$ ratios
(Xu \& Sulentic 1991).  Even among sources from the same subgroups in
this study there is a wide range in star formation activity indicators
(such as FIR/B).  Arp~244, one of the 'advance mergers', shows FIR/B
lower than the 'less disturbed systems' and one of the lowest
$L_{fir}/M_{gas}$ ratios in interacting systems (Gao et al. 1998).  One
explanation for the large scatter in star formation indicators involves
variations in galaxy gas content. Galaxies with less gas have less fuel
for starbursts which is consistent with previous results (Sulentic
1989; Xu \& Sulentic 1991) which found that SS pairs show higher FIR
emission than normal galaxies but little FIR emission from pairs of
early type galaxies (E/S0) which are gas poor. Differences in gas
content between CLO SS pairs with strong and weak star formation
enhancement will be an interesting area of study for future H~I and CO
surveys of galaxy pairs. A surprising lack of H~I gas depletion was
found in a statistical study of an FIR enhanced sample of E+S pairs
(Zasov \& Sulentic 1994).

Even within a sample of gas rich pairs significant differences in
their star formation rates are found. This is true even for galaxies with the
same gas surface density where the ``star formation efficiency'' can
differ by as much as an order of magnitude (see, e.g., Fig.~2 of
Solomon \& Sage 1988).  Much of this scatter may reflect the episodic
nature of starbursts. Several physical mechanisms may contribute to
this episodic behavior.  Feed-back from massive star formation (such as
supernovae explosions and stellar winds) may quench a burst after a few
10$^7$ yr (Kr\"ugel \& Tutukov 1993). This effectively breaks the
interaction induced star formation into pulses which is confirmed by
observations: All starburst durations derived from observation (Rieke
et al. 1980; Gehrz et al. 1983; Bernl\"ohr 1993a) are on the order of a
few times 10$^7$ years even though interactions typically last for
several 10$^8$ years (e.g. Hibbard \& Yun 1999). When an interacting
galaxy is observed in the 'off' stage, only a 'post-starburst' is seen
(examples can be found in Bernl\"ohr 1993a). A 'post-starburst', with
most of the OB stars already gone, will have a much reduced effect on
star formation indicators (e.g. FIR and H$\alpha$ emissions). 

As demonstrated by the simulations of Noguchi (1991), the periodic
orbital motion which swings the two
galaxies back and forth relative to each other several times
before they eventually merge, will
induce sharp peaks (corresponding to the passages of periapses) in the
cloud-cloud collision rate which in turn may also cause pulsational star
formation (see also Olson \& Kwan 1991ab).

The episodic nature of starbursts could be a key to understanding why
only a single component in many pairs shows enhanced star formation
(Joseph et al. 1984). Since the star formation in the members of a pair
may well be unsynchronized (Bernl\"ohr 1993a), there is a good chance
for one of them to be in the 'on' phase while the other is  'off'.  At
the same time there is a significant probability that both components
will be 'on' (Lutz 1992; Surace et al. 1993). 
This suggests that the duration of the 'on'
and 'off' phases must be comparable (a few 10$^7$ yrs).

Another reason for the scatter in star formation activity could be the
interaction geometry. As shown in the pioneering work of
Toomre \& Toomre (1972), retrograde and high inclination encounters cause
much less distortion than direct (prograde) and low inclination
encounters. If, as suggested,  the 'less disturbed systems' have
unfavorable interaction geometries the lower star formation activity
might result from this rather than from an earlier merger evolutionary
stage.  However the evidence that the galaxies in Arp~244 (lowest FIR/B
in our sample) show prograde rotation (van der Hulst
1979) while IC~694 in Arp~299 (highest FIR/B
in our sample) shows retrograde rotation (Hibbard \& Yun
1999), argues against this possibility (see also
Keel 1993; Lutz 1992). This is not meant to imply that orbital geometry
plays no role in the star formation enhancement process. It suggests
only that this role is likely to be complex. For example an unfavorable
orbital geometry may preserve most of the original gas in a galaxy
until a very late stage in the merger process, similar to the role
played by a massive bulge in the simulation of Mihos \&
Hernquist (1996; see also Evans et al. 2000), making a ULIRG-like
starburst more possible (see, e.g., the simulations of Noguchi 1991).
 
\subsection{Galaxy Pairs and Mergers} 
Throughout this discussion we have adopted a scenario where galaxy
pairs merge after a few close encounters, as implied by models (e.g.
Barnes \& Hernquist 1996, 1998). Recent studies on the cosmic
evolution of merger rate using HST data
(Wu \& Keel 1998; Le F\`evre et al. 2000)
also hint that the time scale of mergers is much shorter than
the Hubble time.
However, pair merger time scales as
short as $\sim 1$ Gyr require an explanation for the large number of
isolated binary galaxies ($\sim 10\%$ of all galaxies, Xu \& Sulentic
1991), the majority of which must have been gravitationally
bound systems for $\sim$10 Gyr
(Chatterjee 1987). They must also account for the rarity of candidate
(early-type) merger postcursors found in the the same environments as
these large pair (and compact group) populations (Sulentic \& Rabaca
1994). 

A possible solution for this dilemma is that today's galaxy pairs are
evolved from galaxy groups.  Through coalescence, within the context of
hierarchical galaxy formation (White 1997), two giant galaxies observed
now may be products of a long history of mergers/accretions of smaller
galaxies which were formed as a bound system.  One possibility is that
some mixed pairs (E+S)) represent the last stages in the coallescence
of a compact group (Rampazzo \& Sulentic 1992).  Such a picture may
apply to the two giant galaxies in the Local Group (Milky Way and M31)
which may represent the early stages in the formation of a CLO SS pair
(e.g., Irion 2000). There is evidence that the Milky Way is still
growing by accretion of  smaller satellites, with the Sagittarius dwarf
galaxy being the current ``meal'' (Buser 2000).  Given the recent
discoveries closely linking galaxy evolution and interactions (Williams
et al. 1996; Williams et al. 1998; Thompson et al. 1999; Elbaz et al.
1999; Puget et al. 1999; Blain et al. 1999), any theory for the
formation/evolution of galaxy pairs must play an integral part in the
formation and evolution of galaxies in general.

\section{Conclusion}
We present MIR imaging and spectroscopic observations for a well
defined sample of eight closely interacting pairs of spiral galaxies
with overlapping disks and enhanced FIR emission. Our goal was to study
the star formation distribution in these pairs with special emphasis on
the importance of overlap starbursts.  We identified three possible
subgroups in our sample according to star formation morphology:
\begin{description}
\item{(1)} advanced mergers; 
\item{(2)} severely disturbed systems;
\item{(3)} less disturbed systems. 
\end{description}
Overlap region starbursts are detected in all of the five pairs of
subgroups (1) and (2), suggesting that they are a common property of
colliding systems. On the other hand, except for Arp~244, the 'overlap
starburst' is less intense than the nuclear starbursts in such pairs.
Star formation pairs of subgroup (3) is often more widely distributed
in the disks of both components with no evidence for overlap
starbursts. These systems also show a smaller FIR enhancement, implying
weaker star formation and suggesting that they may be in earlier
interaction stages where direct disk-disk collisions have not yet
occurred.  Only one pair (Arp~157) is detected in our ISO SWS
observations.  The F([Ne~II~12.81$\mu m$])/F(FIR) ratio in the NGC~520b
component of Arp~157 is $0.1$, comparable to values for other luminous
infrared galaxies.


\vskip2cm 
C.X. thanks Marc Sauvage for valuable assistance in reducing the ISO~CAM data,
and Eckard Sturm and Alberto Noriega-Crespo for helping reducing
the SWS data. 
This work was supported by NASA grant for ISO data analysis.
This research has made use of the NASA/IPAC Extragalactic Database (NED) 
which is operated by the Jet Propulsion Laboratory, California Institute of 
Technology, under contract with the National Aeronautics and
Space Administration. 
C.X., Y.G., J.M., and N.L. were supported by the Jet Propulsion Laboratory,
California Institute of Technology, under contract with NASA. 


\noindent{\bf Table 1. CLO SS Pairs Sample}
\nopagebreak

\hskip-3truecm\begin{tabular}{lcccccccccc}\hline
(1)      &  (2)         & (3)          & (4)& (5)   & (6)   & (7) & (8) & (9) & (10) & (11) \\
Name     & R.A.  & DEC. & B  & d & T   & z   & SEP & $L_{fir}\over L_B$ & $f_{60\mu m}\over f_{100\mu m}$ & L$_{fir}$\\
         & (J2000) & (J2000) & mag & $'$ (kpc) & &  km/s & $'$ (kpc) & & & 10$^{10}$L$_\sun$ \\ \hline
Arp~157$^\star$: & & & & & & & 0.65 (5.6) & 4.37 & 0.68 & 4.56 \\
~~NGC~520a & 01h24m32.8s & +03d47m56s & 12.59 & 1.66 (14.4) & Sc &  2105 & && \\  
~~NGC~520b & 01h24m34.9s & +03d47m30s & 13.55 & 1.14  (9.9) & Sm &  2360 & && \\ 
Arp~276:  & & & & & & & 1.06 (16.7) & 1.74 & 0.33 & 2.19 \\
~~NGC~935 & 02h28m11.1s & +19d35m58s & 13.74 & 1.76 (27.7) & Sb &  4142 & && \\
~~IC~1801 & 02h28m12.7s & +19d35m00s & 14.65 & 1.19 (18.7) & Sb &  3970 & && \\
Arp~299: & & & & & & & 0.38 (4.6) & 9.12 & 0.99 & 26.6\\
~~NGC~3690 & 11h28m31.0s & +58d33m41s & 12.72 & 1.30 (15.8) & Sc &  3132 & && \\
~~IC~694   & 11h28m33.5s & +58d33m47s & 12.49 & 1.47 (17.8) & Sc &  3121 & && \\ 
Arp~244:  & & & & & & & 1.21 (7.7) & 1.15 & 0.59 & 3.86 \\
~~NGC~4038  & 12h01m52.8s & -18d51m54s & 10.91 & 5.25 (33.4) & Sm &  1642 & && \\
~~NGC~4039  & 12h01m55.2s & -18d53m06s & 11.05 & 3.09 (19.7) & Sm &  1641 & && \\ 
KPG~347:  & & & & & & & 1.30 (11.4) & 1.20 & 0.36 & 3.18\\
~~NGC~4567  & 12h36m32.7s & +11d15m28s & 12.02 & 2.47 (21.7) & Sb &  2274 & && \\
~~NGC~4568  & 12h36m34.7s & +11d14m15s & 11.99 & 3.50 (30.7) & Sc &  2255 & && \\ 
KPG~426: & & & & & & & 0.88 (26.3) & 1.55 & 0.38 & 4.01 \\
~~UGC~9376a & 14h33m46.8s & +40d04m52s & 14.80 & 1.35 (40.3) & Sb &  7616 & && \\
~~UGC~9376b & 14h33m48.4s & +40d05m39s & 14.56 & 1.52 (45.3) & Sa &  7764 & && \\
Arp~81$^\star$:   & & & & & & & 0.73 (17.9) & 3.39 & 0.55 & 8.36 \\
~~NGC~6621 & 18h12m54.7s & +68d21m49s & 14.39 & 0.96 (23.6) & Sb &  6210 & && \\ 
~~NGC~6622 & 18h13m00.2s & +68d21m12s & 14.23 & 1.00 (24.6) & Sa &  6466 & && \\ 
Arp~278$^\star$:  & & & & & & & 0.84 (15.0) & 6.61  & 0.46 & 4.57 \\
~~NGC~7253a & 22h19m26.2s & +29d23m55s & 15.03 & 1.46 (26.1) & Sm &  4718 & && \\ 
~~NGC~7253b & 22h19m31.3s & +29d23m25s & 14.95 & 1.42 (25.4) & Sm &  4493 & && \\ 
\hline
\end{tabular}
\oneskip
\oneskip

\begin{description}
\item{Col.(1):} Names of galaxy pairs and of pair components.  
Pairs with ISO~SWS observations (see Table 3) are marked by $\star$.
\item{Col.(2):} 2000 epoch right ascencion.
\item{Col.(3):} 2000 epoch declination. 
\item{Col.(4):} Blue magnitude.
\item{Col.(5):} Major axis.
\item{Col.(6):} Hubble type.
\item{Col.(7):} Redshift.
\item{Col.(8):} Component separation.
\item{Col.(9):} FIR-to-blue luminosity ratio.
\item{Col.(10):}  $f_{60\mu m}\over f_{100\mu m}$ color ratio.
\item{Col.(11):} Integrated FIR (40--120 $\mu m$) luminosity calculated from
IRAS 60$\mu m$ and 100$\mu m$ fluxes (Helou et al. 1988) and the mean redshift
of the two components (H$_0$=75 km/s Mpc$^{-1}$).
\end{description}

\clearpage

\noindent{{\bf Table 2. ISO~CAM Observations of CLO SS Pairs}}

\hskip-2truecm\begin{tabular}{lccccccccc}\hline
Name     & Obs.     & Filter    & $\lambda$ & $\lambda/\delta\lambda$ & pixel & samp. & M$\times$N & map  & rms noise   \\
         & date     &           & ($\mu m$) &                         & size  & step  &            & size & (1$\sigma$) \\ 
\hline
Arp~157  &12.29.96 & LW7 & 9.62 & 4 & $3''\times 3''$ &  $45''\times 45''$ & $4\times 4$ & $3.8'\times 3.8'$ & 0.09 mJy/pix \\
         &12.29.96 & LW3 & 15.0 & 3 & $3''\times 3''$ &  $45''\times 45''$ & $4\times 4$ & $3.8'\times 3.8'$ & 0.09 mJy/pix \\ 
Arp~276  & 8.26.96 & LW7 & 9.62 & 4 & $6''\times 6''$ &  $60''\times 60''$ & $3\times 3$ & $5.0'\times 5.0'$ & 0.11 mJy/pix \\
         & 8.26.96 & LW3 & 15.0 & 3 & $6''\times 6''$ &  $60''\times 60''$ & $3\times 3$ & $5.0'\times 5.0'$ & 0.15 mJy/pix \\
Arp~299  & 4.27.96 & LW7 & 9.62 & 4 & $3''\times 3''$ &  $45''\times 45''$ & $4\times 4$ & $3.8'\times 3.8'$ & 0.09 mJy/pix \\
         & 4.27.96 & LW9 & 15.0 & 8 & $3''\times 3''$ &  $45''\times 45''$ & $4\times 4$ & $3.8'\times 3.8'$ & 0.20 mJy/pix \\
Arp~244  & 7.16.96 & LW7 & 9.62 & 4 & $6''\times 6''$ &  $60''\times 60''$ & $4\times 4$ & $6.0'\times 6.0'$ & 0.21 mJy/pix \\
         & 7.16.96 & LW3 & 15.0 & 3 & $6''\times 6''$ &  $60''\times 60''$ & $4\times 4$ & $6.0'\times 6.0'$ & 0.17 mJy/pix \\
Arp~244-02 & 7.28.96 & LW3 & 15.0 & 3 & $6''\times 6''$ &  $6''\times 6''$ & $6\times 6$ & $3.5'\times 3.5'$ & 0.06 mJy/pix \\
KPG~347  &  7.4.96 & LW7 & 9.62 & 4 & $6''\times 6''$ &  $90''\times 90''$ & $3\times 3$ & $6.0'\times 6.0'$ & 0.47 mJy/pix \\
KPG~426  & 8.11.96 & LW7 & 9.62 & 4 & $6''\times 6''$ &  $60''\times 60''$ & $3\times 3$ & $5.0'\times 5.0'$ & 0.11 mJy/pix \\
         & 8.11.96 & LW3 & 15.0 & 3 & $6''\times 6''$ &  $60''\times 60''$ & $3\times 3$ & $5.0'\times 5.0'$ & 0.11 mJy/pix \\
Arp~81   & 8.19.96 & LW7 & 9.62 & 4 & $3''\times 3''$ &  $45''\times 45''$ & $4\times 4$ & $3.8'\times 3.8'$ & 0.10 mJy/pix \\
         & 8.19.96 & LW3 & 15.0 & 3 & $3''\times 3''$ &  $45''\times 45''$ & $4\times 4$ & $3.8'\times 3.8'$ & 0.09 mJy/pix \\
Arp~278  &11.16.96 & LW7 & 9.62 & 4 & $6''\times 6''$ &  $60''\times 60''$ & $3\times 3$ & $5.0'\times 5.0'$ & 0.10 mJy/pix \\
         &11.16.96 & LW3 & 15.0 & 3 & $6''\times 6''$ &  $60''\times 60''$ & $3\times 3$ & $5.0'\times 5.0'$ & 0.12 mJy/pix \\
  \hline

\end{tabular}

\clearpage

\noindent{{\bf Table 3.1. ISO~SWS Observations of CLO SS Pairs}}\hfill

\begin{tabular}{lcccc}\hline
Name    & Pointing$^\dagger$ & R.A. (2000) & Dec. (2000) & Obs. date \\ \hline
Arp~157 & N1 & 1h24m34.9s & 3d $47'$ $29.8''$ & 6.24.97    \\
        & N2 & 1h24m32.8s & 3d $47'$ $55.9''$ & 6.24.97    \\
        & OV & 1h24m33.9s & 3d $47'$ $42.8''$ & 6.24.97    \\
Arp~81  & N1 & 18h12m55.9s & 68d $21'$ $50.1''$ & 4.28.97  \\
        & N2 & 18h13m00.1s & 68d $21'$ $12.3''$ & 4.28.97  \\
        & OV & 18h12m58.1s & 68d $21'$ $31.2''$ & 5.29.97  \\
Arp~278 & N1 & 22h19m26.2s & 29d $23'$ $53.2''$ & 5.9.97   \\
        & N2 & 22h19m28.9s & 29d $23'$ $11.3''$ & 5.9.97   \\ \hline
\end{tabular}
\oneskip
\begin{description}
\item{$^\dagger$} Centers of ISO~SWS pointings:
   \begin{description}
     \item{N1 ---} Nucleus of component 1 of the galaxy pair.
     \item{N2 ---} Nucleus of component 2 of the galaxy pair.
     \item{OV ---} Overlap region of the two components. 
   \end{description}
\end{description}
\oneskip

\noindent{{\bf Table 3.2. Parameters of ISO~SWS Observations}}\hfill

\begin{tabular}{lccccc}\hline
           & $\lambda$ & SWS- & Aperture        & Resolution               & Sensitivity \\
 Line      & ($\mu m$) & band & ($''\times ''$) & ($\lambda/\delta\lambda$)& ($3\sigma$, Jy) \\ \hline
Br$_\beta$  &   2.63   & SW-1B & $14\times 20$ & 1470--1750 & 0.057 \\ 
Br$_\alpha$ &   4.05   & SW-2A & $14\times 20$ & 1540--2130 & 0.29  \\ 
$[$Ne~II$]$     &   12.81  & LW-3A & $14\times 27$ & 1250--1760 & 0.80  \\ 
H$_2$ S(1)  &   17.03  & LW-3C & $14\times 27$ & 1760--2380 & 0.69  \\ 
\hline
\end{tabular}

\noindent{{\bf Table 4. ISO~CAM Fluxes of CLO SS Pairs Sample}}
\nopagebreak

\begin{tabular}{lll}\hline
(1) & (2) & (3) \\
Name     & f$_{9.7\mu}$ & f$_{15\mu}$ \\
         & (mJy) & (mJy) \\ \hline
Arp~157: & 358  & 766 \\
~~NGC~520a & 32 & 49 \\
~~NGC~520b & 327 & 715 \\
Arp~276:  &   241 & 300 \\ 
~~NGC~935 & 183 & 228 \\
~~IC~1801 & 63 & 72 \\
Arp~299: & 1699 & 5977 \\
~~NGC~3690 & 1071 & 3494 \\
~~IC~694   & 550 & 2158 \\
Arp~244:  & 1227 & 2124 \\
~~NGC~4038  & 489 & 700 \\
~~NGC~4039  & 738 & 1427 \\
KPG~347$^\dagger$:  & 914 & 1269 \\
~~NGC~4567$^\dagger$  & 124 & 252 \\
~~NGC~4568$^\dagger$  & 790 & 1017 \\
KPG~426: & 84 & 110 \\
~~UGC~9376a & 32 & 41 \\
~~UGC~9376b & 52 & 66 \\
Arp~81:   & 171 & 270 \\ 
~~NGC~6621 & 129 & 216 \\
~~NGC~6622 & 22 & 15 \\
Arp~278:  & 178 & 253 \\
~~NGC~7253a & 126 & 200 \\
~~NGC~7253b & 44 & 49 \\
\hline
\end{tabular}

\oneskip
$^\dagger$ The 15$\mu m$ data are taken from Bosselli et al. 1998.
\oneskip

\begin{description}
\item{Col.(1):} Names of galaxy pairs and of pair components.  
\item{Col.(2):} Flux density at 9.7$\mu m$. 
The uncertainty is $\sim 15\%$, dominated by the calibration error.
\item{Col.(3):} Flux density at 15$\mu m$. 
The uncertainty is $\sim 15\%$, dominated by the calibration error.
\end{description}

\clearpage

\noindent{{\bf Table 5. SWS line emission of CLO SS Pairs}}

\begin{tabular}{cccccc}\hline
        &       &\multicolumn{4}{c}{Flux ($10^{-20}$ W/cm$^2$)} \\ \cline{3-6} 
Name    & Pointing & Br$_\beta$ & Br$_\alpha$ & [Ne~II] & H$_2$ S(1) \\ 
  & & $2.63\mu m$ & $ 4.05\mu m$ & $ 12.81\mu m$ & $ 17.03\mu m$ \\ 
\hline 
&&&&& \\
Arp~157 & N1 & $<2.3 $ & $<2.2$ & 19.3 ($\pm 1.5$) & 2.60 ($\pm 0.34$) \\
        & N2 & $<0.6$ & $<2.2$ & $<1.4$ & $<1.4$ \\
        & OV & $<0.6$ & $<2.2$ & 4.88 ($\pm 0.96$)  & $<1.4$ \\
&&&&& \\
Arp~81  & N1 & $<0.6$ & $<3.0$ & $<1.4$ & $<1.4$ \\
        & N2 & $<2.3 $ & $<2.2$ & $<0.7$ & $<0.5$ \\
        & OV & $<0.6$ & $<3.0$ & $<1.9$ & $<1.2$ \\
&&&&& \\
Arp~278 & N1 & $<0.4$ & $<3.0$ & $<1.2$ & $<0.9$ \\
        & N2 & $<0.5$ & $<2.2$ & $<2.3$ & $<1.2$ \\
&&&&& \\
\hline 
\end{tabular}
\clearpage

\noindent{{\bf Table 6. ISO~CAM Fluxes of Nuclear and
Overlap Starburst Emission in Arp~299 and Arp~244}}

\begin{tabular}{lllll}\hline
&&&& \\
(1)      &  (2)         & (3) & (4) & (5) \\
Name     & R.A.  & DEC.    & f$_{9.7\mu}$ & f$_{15\mu}$ \\
         & (J2000) & (J2000)         & (mJy) & (mJy) \\

&& &&\\
Arp~299: & & &&\\
~~NGC~3690$^\star$     & 11h28m31.0s & +58d33m41s & 613 ($\pm 65$) & 1410 ($\pm 150$)\\
~~IC~694$^\star$       & 11h28m33.5s & +58d33m47s & 221 ($\pm 25)$) & 1044 ($\pm 110$)\\
~~overlap starburst$^\ddagger$ & 11h28m31.0s & +58d33m49s & 123 ($\pm 25$) & 471 ($\pm 70$)\\  
&&&& \\
Arp~244:  & & &&\\
~~NGC~4038$^\star$  & 12h01m52.8s & -18d51m54s  & 74 ($\pm 10$) & 135 ($\pm 20$) \\
~~NGC~4039$^\star$  & 12h01m55.2s & -18d53m06s  & 39 ($\pm 10$) & 76 ($\pm 20$)\\
~~overlap starburst$^{\star}$ & 12h01m54.8s & -18d53d03s & 142 ($\pm 18$) & 359 ($\pm 50$) \\
&&&& \\ \hline

\end{tabular}
\oneskip

\begin{description}
\item{$^\star$} Fluxes estimated from point source extractions.
\item{$^\ddagger$} 
         Fluxes estimated by summing up
         the counts in a region of size$\sim 15''$ after 
         the source associated to the nucleus of NGC~3690 is 
         extracted.
\end{description}

\clearpage


\clearpage

\figcaption[fig1.ps]{
MIR and H$\alpha$ emissions of Arp~299 
and Arp~244. {\bf Arp~299:} (a) Contours of 15$\mu m$ emission on
optical (DSS) image. Contour levels are 2$^n$ (n=1,2,3,...)
$\times \sigma_{15\mu m}$ ($\sigma_{15\mu m} = 22\mu Jy/arcsec^2$). 
Scale: 1$^\prime = 12.1$ kpc.
(b) Grayscale (logarithmic scale) image of
$f_{15\mu m}/f_{9.7\mu m}$ ratio, in the range of 1 (dark) and
10 (bright), overlaid with 15$\mu m$ contours
(levels: $3+3^n$ (n=0,1,2,...) $\times \sigma_{15\mu m}$).
(c) H$\alpha$ contours (levels: $3+3^n$ (n=1,2,3,...) $\times 10^{-17}\;
erg\, sec^{-1}cm^{-2}pix^{-1}$, $pix=0.62^{\prime\prime}$)
on R band images. Note that the overlap starburst (Source C in 
Gehrz et al. 1983) is only $\sim 8^{\prime\prime}$ above the
nucleus of NGC~3690 (the western galaxy). 
{\bf Arp~244:} (d) Contours of 15$\mu m$ emission on
optical (DSS) image. Contour levels are 2$^n$ (n=1,2,3,...)
$\times \sigma_{15\mu m}$ ($\sigma_{15\mu m} = 4.7\mu Jy/arcsec^2$). 
Scale: 1$^\prime = 6.4$ kpc.
(e) Grayscale (logarithmic scale) image of
$f_{15\mu m}/f_{9.7\mu m}$ ratio, in the range of 1 (dark) and
4 (bright), overlaid with 15$\mu m$ contours
(levels: $3+3^n$ (n=0,1,2,...) $\times \sigma_{15\mu m}$).
(f) H$\alpha$ contours (levels: $2+3^n$ (n=1,2,3,...) $\times 10^{-17}\;
erg\, sec^{-1}cm^{-2}pix^{-1}$, $pix=0.62^{\prime\prime}$)
on R band images. Note that the overlap starburst (Source A in 
Vigroux et al. 1996) coincides with the southern peak of the 15$\mu m$   
emission, while the nucleus of NGC~4039 (the southern galaxy) is
$\sim 10^{\prime\prime}$ southwest of the peak. 
}
\vskip1truecm

\figcaption[fig2.ps]{
MIR and H$\alpha$ emissions of Arp~157 
and Arp~81. {\bf Arp~157:} (a) Contours of 15$\mu m$ emission on
optical (DSS) image. Contour levels are 2$^n$ (n=1,2,3,...)
$\times \sigma_{15\mu m}$ ($\sigma_{15\mu m} = 10\mu Jy/arcsec^2$). 
Scale: 1$^\prime = 8.6$ kpc.
(b) Grayscale (logarithmic scale) image of
$f_{15\mu m}/f_{9.7\mu m}$ ratio, in the range of 1 (dark) and
2.5 (bright), overlaid with 15$\mu m$ contours
(levels: $3+3^n$ (n=0,1,2,...) $\times \sigma_{15\mu m}$).
(c) H$\alpha$ contours (levels: $3+3^n$ (n=1,2,3,...) $\times 10^{-17}\;
erg\, sec^{-1}cm^{-2}pix^{-1}$, $pix=0.53^{\prime\prime}$)
on R band images. Note that the H$\alpha$ observation was
nonphotometric, so the calibration is a rough estimate.  
{\bf Arp~81:} (d) Contours of 15$\mu m$ emission on
optical (DSS) image. Contour levels are 2$^n$ (n=1,2,3,...)
$\times \sigma_{15\mu m}$ ($\sigma_{15\mu m} = 10\mu Jy/arcsec^2$). 
Scale: 1$^\prime = 24.6$ kpc.
(e) Grayscale (logarithmic scale) image of
$f_{15\mu m}/f_{9.7\mu m}$ ratio, in the range of 1 (dark) and
3.16 (bright), overlaid with 15$\mu m$ contours
(levels: $3+3^n$ (n=0,1,2,...) $\times \sigma_{15\mu m}$).
(f) H$\alpha$ contours (levels: $1+3^n$ (n=1,2,3,...) $\times 10^{-17}\;
erg\, sec^{-1}cm^{-2}pix^{-1}$, $pix=0.62^{\prime\prime}$)
on R band images.
}
\vskip1truecm

\figcaption[fig3.ps]{
MIR and H$\alpha$ emissions of Arp~278 
and Arp~276. {\bf Arp~278:} (a) Contours of 15$\mu m$ emission on
optical (DSS) image. Contour levels are 2$^n$ (n=1,2,3,...)
$\times \sigma_{15\mu m}$ ($\sigma_{15\mu m} = 3.3\mu Jy/arcsec^2$). 
Scale: 1$^\prime = 17.9$ kpc.
(b) Grayscale (logarithmic scale) image of
$f_{15\mu m}/f_{9.7\mu m}$ ratio, in the range of 1 (dark) and
2 (bright), overlaid with 15$\mu m$ contours
(levels: $3+3^n$ (n=0,1,2,...) $\times \sigma_{15\mu m}$).
(c) H$\alpha$ contours (levels: $1+3^n$ (n=1,2,3,...) $\times 10^{-17}\;
erg\, sec^{-1}cm^{-2}pix^{-1}$, $pix=0.33^{\prime\prime}$)
on R band images.  
{\bf Arp~276:} (d) Contours of 15$\mu m$ emission on
optical (DSS) image. Contour levels are 2$^n$ (n=1,2,3,...)
$\times \sigma_{15\mu m}$ ($\sigma_{15\mu m} = 4.2\mu Jy/arcsec^2$). 
Scale: 1$^\prime = 15.7$ kpc.
(e) Grayscale (logarithmic scale) image of
$f_{15\mu m}/f_{9.7\mu m}$ ratio, in the range of 1 (dark) and
2 (bright), overlaid with 15$\mu m$ contours
(levels: $3+3^n$ (n=0,1,2,...) $\times \sigma_{15\mu m}$).
(f) H$\alpha$ contours (levels: $1+3^n$ (n=1,2,3,...) $\times 10^{-17}\;
erg\, sec^{-1}cm^{-2}pix^{-1}$, $pix=0.53^{\prime\prime}$)
on R band images. Note that the H$\alpha$ observation was
nonphotometric, so the calibration is a rough estimate. 
}
\vskip1truecm

\figcaption[fig4.ps]{
MIR and H$\alpha$ emissions of KPG~347
and KPG~426. {\bf KPG~347:} (a) Contours of 9.7$\mu m$ emission on
optical (DSS) image. Contour levels are 2$^n$ (n=1,2,3,...)
$\times \sigma_{9.7\mu m}$ ($\sigma_{9.7\mu m} = 13\mu Jy/arcsec^2$). 
Scale: 1$^\prime = 8.8$ kpc.
(b) Grayscale (logarithmic scale) image of
$f_{15\mu m}/f_{9.7\mu m}$ ratio, in the range of 1 (dark) and
2 (bright), overlaid with 9.7$\mu m$ contours
(levels: $3+3^n$ (n=0,1,2,...) $\times \sigma_{9.7\mu m}$).
(c) H$\alpha$ contours (levels: $1+3^n$ (n=1,2,3,...) $\times 10^{-17}\;
erg\, sec^{-1}cm^{-2}pix^{-1}$, $pix=0.33^{\prime\prime}$)
on R band images.  
{\bf KPG~426:} (d) Contours of 15$\mu m$ emission on
optical (DSS) image. Contour levels are 2$^n$ (n=1,2,3,...)
$\times \sigma_{15\mu m}$ ($\sigma_{15\mu m} = 3.1\mu Jy/arcsec^2$). 
Scale: 1$^\prime = 30.0$ kpc.
(e) Grayscale (logarithmic scale) image of
$f_{15\mu m}/f_{9.7\mu m}$ ratio, in the range of 1 (dark) and
2 (bright), overlaid with 15$\mu m$ contours
(levels: $3+3^n$ (n=0,1,2,...) $\times \sigma_{15\mu m}$).
(f) H$\alpha$ contours (levels: $1+3^n$ (n=1,2,3,...) $\times 10^{-17}\;
erg\, sec^{-1}cm^{-2}pix^{-1}$, $pix=0.33^{\prime\prime}$)
on R band images. Note that the H$\alpha$ observation was
nonphotometric, so the calibration is a rough estimate. 
}
\vskip1truecm

\figcaption[fig5.ps]{
[NeII~12.81$\mu m$] line emission from the Arp~157-N1 region
(nucleus of NGC~520b).
}
\vskip1truecm

\figcaption[fig6.ps]{
[H$_2$~S(1)~17.03$\mu m$] line emission from the Arp~157-N1 region
(nucleus of NGC~520b).
}
\vskip1truecm

\figcaption[fig7.ps]{
[NeII~12.81$\mu m$] line emission from the Arp~157-OV region (between
nuclei of NGC~520a and NGC~520b).
}



\begin{references}

\reference{aalto97} Aalto, S., Radford, S.J.E., Scoville, N.Z., Sargent, 
A.I. 1997, \apj, 475, L107.

\reference{arp66} Arp, H.C. 1966, ApJS, 14, 1.

\reference{augarde85} Augarde, R., Lequeux, J. 1985, \aap, 147, 273.

\reference{bal83} Balzano, V.A. 1983, \apj, {268}, 602-627 (1983).

\reference{barnes96} Barnes, J., Hernquist, L. 
1996, \apj, 471, 115.

\reference{barnes98} Barnes, J., Hernquist, L. 
1998, \apj, 495, 187.

\reference{Ber93bh} Bernl\"ohr, K. 1993a, \aap, 268, 25.

\reference{Ber93} Bernl\"ohr, K. 1993b, \aap, 270, 20.

\reference{bla99} Blain, A.W., Smail, I., Ivison, R.J., Kneib, J.-P.
1999, astro-ph/9908111.

\reference{Bos98} Boselli, A., Lequeux, J., Sauvage, M., et al. 1998, \aap, 335, 538.

\reference{Br99} Bryant, M.B., Scoville, N.Z. 1999, \aj, 117, 2632.

\reference{bus98} Bushouse, H.A., Telesco, C.M., Werner, M.W. 1998, \aj, 115, 938.

\reference{bus90} Bushouse, H.A., Werner, M.W. 1990, \apj, 359, 72.

\reference{bus00} Buser, R. 2000, Science, 287, 69.

\reference{casoli99} Casoli, F., Willaime, M.-C., Viallefond, F., Gerin, M. 1999, \aap, 346, 663.

\reference{cayatte94} Cayatte, V., Kotanyi, C., Balkowski, C., 
van Gorkom, J.H. 1994, \aj, 107, 1003

\reference{cesarsky96} Cesarsky, C.J., Abergel, A., Agn\'ese, P.,
et al. 1996, \aap, {315}, L32.

\reference{Chatterjee87} Chatterjee, T.K. 1987, Ap\&SS, 135, 131.

\reference{Clements96} Clements. D.L., Sutherland, W.J., McMahon, R.G., 
Saunders, W. 1996, MNRAS, 279, 477.

\reference{ClementsB96} Clements. D.L., Baker, A.C. 1996, \aap, 314, L5.

\reference{con82} Condon, J.J., Condon, M.A., Gisler, G., Puschell, J.J. 1982, ApJ, 252, 102.

\reference{con93} Condon, J.J., Helou, G., Sanders, D.B., Soifer, B.T. 1993, \aj, 105, 1730.

\reference{con96} Condon, J.J., Helou, G., Sanders, D.B., Soifer, B.T. 1996, ApJS, 103, 81.

\reference{corbelli91} Corbelli, E., Salpeter, E.E., Dickey , J.M. 1991,
\apj, 347, 49.

\reference{cut85} Cutri, R., McAlary, C.W. 1985, ApJ, 296, 90.

\reference{d90} D\'esert, F.-X., Boulanger, F., Puget, J.-L.
1990, \aap, 237, 215.

\reference{dopita85} Dopita, M.A. 1985, \apj, 295, L5.

\reference{downes98} Downes, D., Solomon, P.M. 1998, \apj, 507, 615.

\reference{elb99} Elbaz, D., Cesarsky, C.J., Fadda, D., et al. 
1999, A\&A, 351, L37.

\reference{evans98} Evans, R., Harper, A., Helou, G. 1998, in {\it Extragalactic Astronomy in Infrared}, Proceedings of the XVIIth Moriond Astrophysical Meetings, eds. G.A. Mamon, T.X. Thuan and J.T. Thanh Van; p143.

\reference{evans00} Evans, A.S., Surace, J.A., Mazzarella, J.M. 2000, 
\apj, 529, L85.

\reference{fabbiano97} Fabbiano, G., Schweizer, F., Mackie, G. 1997,
\apj, 478, 542.

\reference{gaoetal98}
        Gao, Y., Gruendl, R.A., Lo, K.Y., Lee, S.-W., Hwang, C.Y. 
        1998, AAS, 192, \#69.04. 

\reference{gao99} Gao, Y., Solomon, P.M. 1999, \apj, 512, L99.

\reference{gallais99} Gallais, P., Laurent, O., Charmandaris, V., et al. 1999,
 Proceedings of ''The Universe a seen by ISO'' meeting,
ESA SP-427, March 1999, p. 881. 

\reference{geh83} Gehrz, R.D., Sramek, R.A., Weedman, D.W. 1983,
\apj, {267}, 551.

\reference{gen98} Genzel, R., Lutz, D., Sturm, E., et al. 1998,
\apj, 495, 1998.

\reference{gold78} Goldreich, P., Tremaine, S. 1978, ApJ, 222, 850.

\reference{gruendl98}
        Gruendl R.A., Gao, Y., Lo, K.Y., Hwang, C.-Y., Lee, S.-W. 
        1998, AAS, 192, \#69.05.

\reference{haas00} Haas, M., Klaas, U., Coulson, I. et al. 2000, 
\aap, submitted.

\reference{helou98} Helou, G., Khan, I.R., Malek, L., Beohmer, L. 
1988, \apjs{68}, 151.

\reference{hib99} Hibbard, J.E., van Gorkom, J.H. 1996, \aj, 111, 655.

\reference{hib99} Hibbard, J.E., Yun, M.S. 1999, \aj, 118, 162. 

\reference{hum86} Hummel, E., van der Hulst, J.M. 1986, \aap, 155, 151.

\reference{hwang99} Hwang, Y.H., Lo, K.Y., Gao, Y., Gruendl, R.A., 
Lu, N.Y. 1999, \apj 511, L17.

\reference{hunter98} Hunter, E.A., Elmegren, B.G., Baker, A.L. 1998,
\apj, 493, 595.

\reference{irion00} Irion, R. 2000, Science, 287, 62.

\reference{Jerrett99} Jerry, T.H., Helou, G., Van Buren, D., Valjavec, E. 1999, \apj, 
118, 312.

\reference{Joness88} Jones, B., Stein, W.A. 1989, \aj, 98, 1557.

\reference{jog92} Jog, C.J., Solomon, P.M. 1992, \apj, {387}, 152.

\reference{jos84} Joseph, R.D., Meikle, W.P.S., Robertson, N.A.,
Wright, G.S. 1984, MNRAS, {209}, 111.

\reference{kar72} Karachentsev, I. 1972, Comm. Spec. Ap. Obs., USSR, 7, 1.

\reference{keel93} Keel, W.C. 1993, \aj, 106, 1771.


\reference{ken89}  Kennicutt, R.C. 1989, \apj, 344, 685.

\reference{ken98}  Kennicutt, R.C. 1998, \apj, 498, 541.

\reference{ken87}  Kennicutt, R.C. Keel, W.C., van der Hulst, J.M., et al.
1987, \aj, 93, 1011. 


\reference{kes96} Kessler, M.F., Steinz, J.A., Anderegg, M.E.
et al. 1996, \aap, {315}, L27.

\reference{krugel93} Kr\"ugel, E., Tutukov, A.V. 1993, \aap, 275, 416.

\reference{kun96} Kunze, D., Rigopoulou, D., Lutz, D. et al. 1996; \aap, {315}, L101.

\reference{lar78} Larson, R.B., Tinsley, B.M. 1978, \apj, {219}, 46.

\reference{leFevre} Le F\`evre, O., Abraham, R., Lilly, R.S., et al. 
2000, MNRAS, 331, 565.

\reference{lon84} Lonsdale, C.J., Persson, S.E., Matthews, K.
1984, \apj, {287}, 95.

\reference{lutz92} Lutz, D. 1992, \aap, 259, 426.

\reference{lutz98} 
 Lutz, D., Spoon, H. W. W., Rigopoulou, D., et al. 1998, \apj, 505, L103.

\reference{mazz91} 
Mazzarella, J.M., Gaume, R.A., Soifer, B.T., et al. 1991, \aj, 102, 1241.

\reference{melnick90} 
 Melnick J.,  Mirabel, I.F. 1990, \aap, 231, 19.

\reference{mihos98} Mihos, J.C. Bothun, G.D.  
1998, \apj, {500}, 619.

\reference{mih93} Mihos, J.C. Bothun, G.D., Richstone, D.O.
1993, \apj, {418}, 82.

\reference{mih96} Mihos, J.C., Hernquist, L. 1996,
\apj, {464}, 641.

\reference{mir98} Mirabel, I.F., 
Vigroux, L., Charmandaris, V., et al. 1998, \aap, 333, L1.

\reference{mir92} Mirabel, I.F., Dottori, H., Lutz, D. 
1992, \aap, {256}, L19.

\reference{mir92} Murphy, T., Armus, L., Mathews, K, Soifer, B.T. et al.
1996, AJ, 111, 1025.

\reference{nik98} Nikola, T., Genzel, R., Herrmann, F., et al. 1998, \apj, 504, 749.

\reference{nog88} Noguchi, M.  1988, \aap, 203, 259.

\reference{nog91} Noguchi, M.  1991, MNRAS, 251, 360.

\reference{nog86} Noguchi, M., Ishibashi, S. 1986, MNRAS, {219}, 305.

\reference{ol90a} Olson, K.M., Kwan, J. 1990a, \apj, 349, 480.

\reference{ol90b} Olson, K.M., Kwan, J. 1990b, \apj, 361, 426.

\reference{pie97} Pietsch, W., et al. \aap, 322, 89.
 
\reference{pug99} Puget, J-L., Lagache, G., Clements, D.L, et al. 1999,
\aap 345, 29.


\reference{rampazzo92} Rampazzo, R., Sulentic, J. W. 1992, \aap, 259, 43.

\reference{rie80} Rieke, G.H., Lebofsky, M.J., Thompson, R.I.,
Low, F.J., Tokunaga, A.T. 1980 \apj, {238}, 24.

\reference{rub70} Rubin, V.C., Ford, W.K., D'Odorico, S. 1970, \apj, 160, 801.

\reference{sanm96} Sanders, D.B., Mirabel, I.F. 1996, ARA\&A, 34, 749.

\reference{san88a} Sanders, D.B, Soifer, B.T., Elias, J.H., Madore, B.F., 
Matthews, K., Neugebauer, G., Scoville, N.Z. 1988a, ApJ, 325, L74.

\reference{san88b} Sanders, D.B., Scoville, N.Z., Sargent, A.I., 
Soifer, B.T. 1988b, ApJ, 324, L55.

\reference{san91} Sanders, D.B., Scoville, N.Z., Soifer, B.T. 1991, \apj, 
370, 158.

\reference{sargent91} Sargent, A.I., Scoville, N.Z. 1991, \apj, 366, L1.

\reference{sau96} Sauvage, M., Blommaert, J., Boulanger, F.
et al. 1996, \aap, {315}, L89.

\reference{sch90} Schombert, J.M., Wallin, J.F., Struck-Marcell, C.
1990, \aj, {99}, 497.

\reference{sch78} Schweizer, F. 1978, in IAU Symp. 77, p279.

\reference{scoville94} Scoville, N., Hibbard, J.E., Yun, M.S., 
van Gorkom, J.H. 1994, in ``Mass-Transfer Induced Activity in Galaxies'',
ed. I. Shlosman (Cambridge: Cambridge Univ. Press), p191.

\reference{scoville86} Scoville, N., Sanders, D.B., Sargent, A.I., 
Soifer, B.T., Scott, S.L., Lo, K.Y. 1986, \apj, 324, L55.

\reference{soif87} Soifer, B.T., Sanders, D.B., Madore, B.F.,
Neugebauer, G. et al. 1987, ApJ, 320, 238.

\reference{solomon89} Solomon, P.M., Sage, L. 1988, \apj, 334, 613.

\reference{solomon97} Solomon, P.M., Downes, D., Radford, S.J.E., 
Barrett, J.W. 1997, \apj, 478, 144.

\reference{sta91} Stanford, S.A. 1991, \apj, 381, 409. 

\reference{stab90} Stanford, S.A., Balcells, M. 1990, \apj, 355, 59. 

\reference{stab91} Stanford, S.A., Balcells, M. 1991, \apj, 370, 118. 

\reference{sta90} Stanford, S.A.., Sargent, A.I., 
Sanders, D.B., Scoville, N.Z. 1990, \apj, 349, 492. 

\reference{struck97} Struck, C. 1997, \apjs, 113, 269.

\reference{sulentic89} Sulentic, J. 1989, \aj, 98, 2066.

\reference{sulentic00} Sulentic, J.W. et al. 2000, in ``Cosmic Evolution and Galaxy Formation'', eds. J. Franco et al. (PASP), in press.
 
\reference{sulentic94} Sulentic, J.W., Raba\c ca, C. 1994, \apj, 429, 531. 

\reference{surace93} Surace, J.A., Mazzarella, J., Soifer, B.T., Wehrle, A.E.
1993, AJ, 105, 864.

\reference{surace99} Surace, J.A., Sanders, D.B. 1999, \apj, 512, 162.

\reference{tele85} Telesco, C.M., Decher, R., Gatley, I. 
1985, \apj, {299}, 896.

\reference{tele88} Telesco, C.M., Wolstebcroft, R., Done, C. 1988, 
\apj, 329, 174.

\reference{tho89} Thomasson, M., Donner, K.J., 
Sundelius, B., et al. 1989, \aap, 211, 25.

\reference{thom99} Thompson, R.I., Storrie-Lombardi, L.J., Weymann,
R.J., Rieke, M.J., Schneider, G., Stobie, E., Lytle, D. 1999, \aj, 117, 17.

\reference{t78} Toomre, A. 1978, in IAU Symp. 79, p109.

\reference{tt72} Toomre, A., Toomre, J. 1972, \apj, 178, 623.

\reference{van98} Vanzi, L., Alonso-Herrero, A., Rieke, G.H. 1998,
\apj, 504, 93.

\reference{vanderhulst79} van der Hulst, J.M. 1979, \aap, 71, 131.

\reference{vig96} Vigroux, L., Mirabel, F., Alti\'eri, B.
et al. 1996, \aap, {315}, L93.

\reference{young88} Young, J.S., Kleinmann, S.G., Allen, L.E. 
1988, \apj, 334, L63.

\reference{young86} Young, J.S., Schloerb, F.P., Kenney, J.P.D., Lord, S.
1986, \apj, 304, 443.

\reference{young89} Young, J.S., Xie, S., Kenney, J.D.P., Rice, W.L.
1989, \apjs, 70, 699.

\reference{wyse87} Wyse, R.F., Silk, J. 1987, \apj, 339, 700.

\reference{whit97} White, S.D.M. 1997, in ``The Evolution of the 
Universe'', eds. G. Borner and S. Gottlober (New York: J. Wiley),  p227.

\reference{whi99} Whitmore, B.C., Zhang, Q., Leitherer, C., et al. 1999, \aj,
118, 155.

\reference{WW91} Wynn-Williams, C.G.; Eales, S.A., Becklin, E.E., et al. 
1991, \apj, 377, 426.
 
\reference{wil96} Williams, S.E. et al. 1996, \aj, 112, 1335.

\reference{wil98} Williams, S.E. et al. 1998, BAAS, 193. 7501.

\reference{wil98} Wu, W., Keel, W.C. 1998, AJ, 116, 1513.

\reference{Xu89} Xu, C., De Zotti, G. 1989, \aap, 225, 12.

\reference{Xu91} Xu, C., Sulentic, J.W. 1991, \apj, 374, 407.

\reference{Xu99} Xu, C., Sulentic, J.W., Tuffs, R. 1999, \apj, 512, 178.

\reference{za94} Zasov, A., Sulentic, J.W. 1994, \apj, 430, 179. 

\end{references}
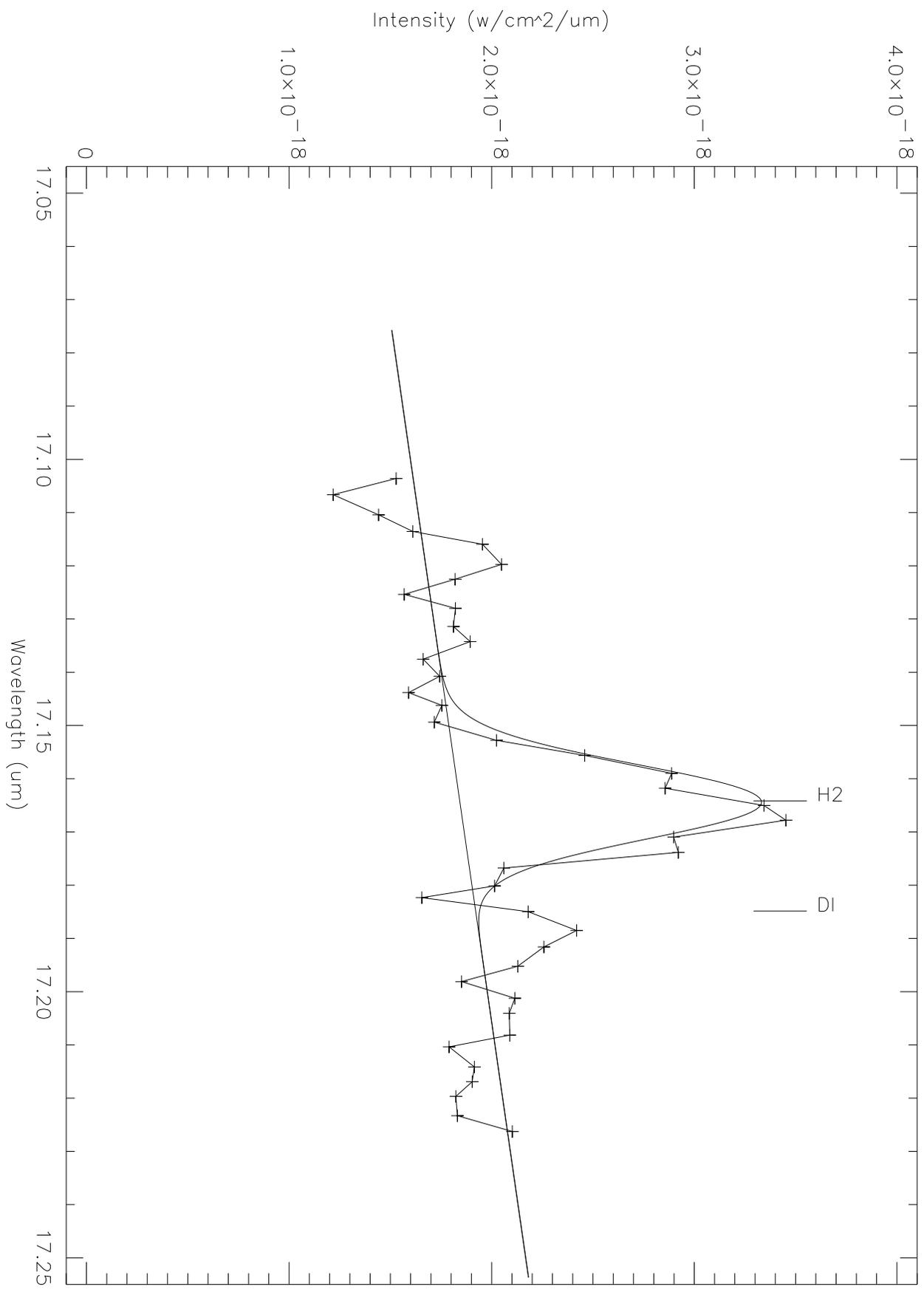
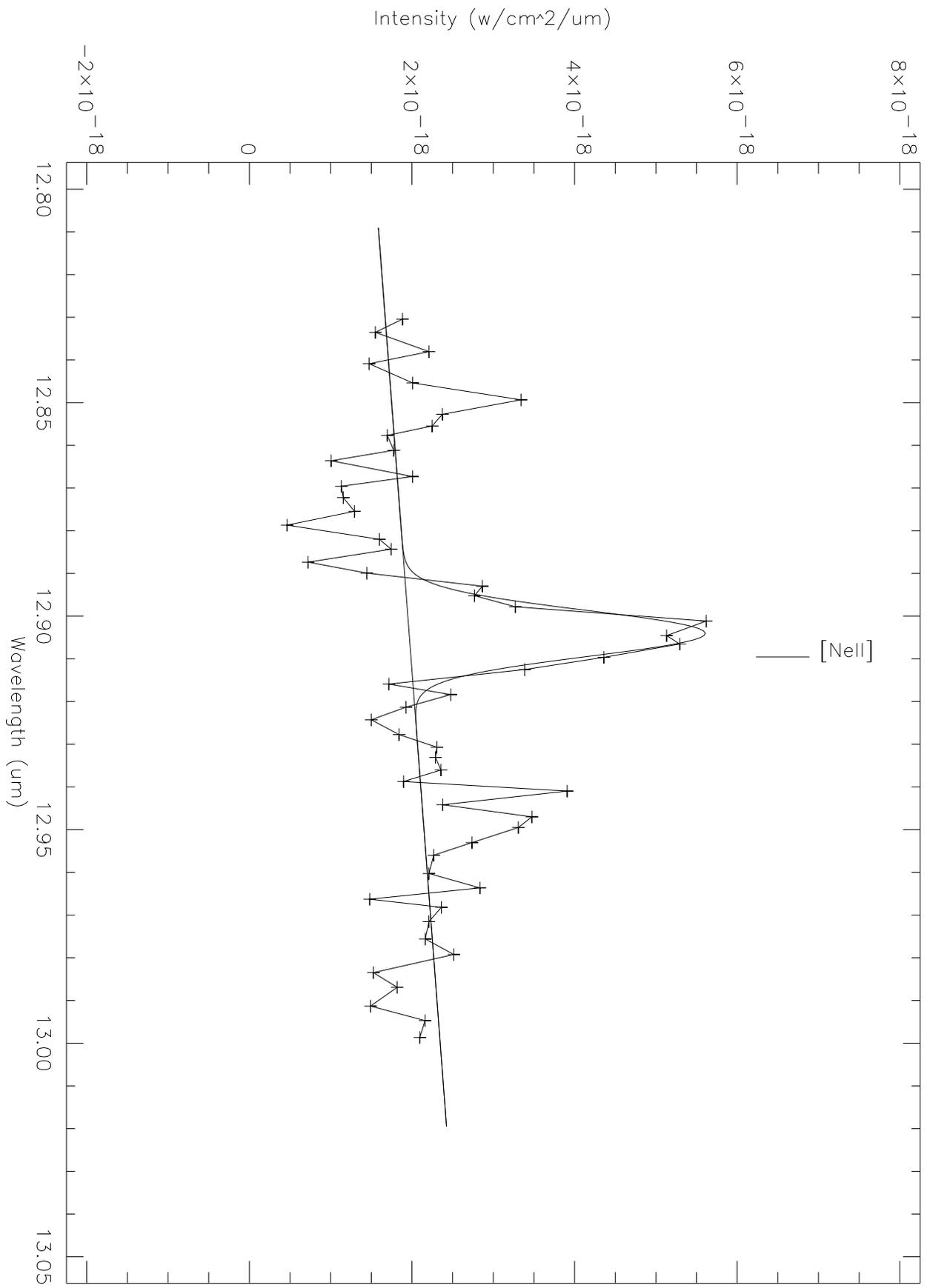
\end{document}